\documentclass[particles,article,accept,pdftex,moreauthors]{Definitions/mdpi} 
\firstpage{1} 
\pubvolume{1}
\issuenum{1}
\articlenumber{0}
\pubyear{2025}
\copyrightyear{2025}
\datereceived{22 October 2025 } 
\daterevised{9 December 2025 } 
\dateaccepted{7 January 2026 } 
\datepublished{ } 
\hreflink{https://doi.org/} 

\Title{Gamma-Ray Burst Polarimetry with the COMCUBE-S CubeSat Swarm---Design and Performance Simulations}

\Author{Nathan Franel 
$^{1,}$*, 
 Vincent Tatischeff $^{1,}$*\orcidC{}, David Murphy $^{2}$, Alexey Ulyanov $^{2}$\orcidD{}, Caimin McKenna $^{2}$\orcidB{},\linebreak   Lorraine Hanlon $^{2}$\orcidA{}, Prerna Baranwal $^{3}$,  
Christophe Beigbeder $^{1}$,
Arnaud Claret $^{4}$,
Ion Cojocari $^{1}$,\linebreak
Nicolas de Séréville $^{1}$,
Nicolas Dosme $^{1}$,
Eric Doumayrou $^{4}$,
Mariya Georgieva $^{1}$,
Clarisse Hamadache $^{1}$,\linebreak
Sally Hankache $^{1}$,
Jimmy Jeglot $^{1}$,
M\'{o}zsi Kiss $^{3}$\orcidH{},
Beng-Yun Ky $^{1}$,
Vincent Lafage $^{1}$\orcidI{},
Philippe Laurent $^{4}$\orcidF{},\linebreak
Christine Le Galliard $^{1}$,
Joseph Mangan $^{1}$,
Aline Meuris $^{4}$\orcidG{},
Mark Pearce $^{3}$\orcidE{},
Jean Peyré $^{1}$,
Arjun Poitaya $^{4}$,\linebreak
Diana Renaud $^{4}$,
Arnaud Saussac $^{1}$,
Varun Varun $^{3}$,
Matias Vecchio $^{1}$ and
Colin Wade $^{2}$}

\AuthorNames{Nathan Franel, Vincent Tatischeff,  David Murphy, Alexey Ulyanov, Caimin McKenna, Lorraine Hanlon, Prerna Baranwal, Christophe Beigbeder, Arnaud Claret, Ion Cojocari, Nicolas de Séréville, Nicolas Dosme, Eric Doumayrou, Mariya Georgieva, Clarisse Hamadache, Sally Hankache, Jimmy Jeglot, M\'{o}zsi Kiss, Beng-Yun Ky, Vincent Lafage, Philippe Laurent, Christine Le Galliard, Joseph Mangan, Aline Meuris,  Mark Pearce, Jean Peyré, Arjun Poitaya, Diana Renaud, Arnaud Saussac, Varun Varun, Matias Vecchio and Colin Wade}

\address{%
$^{1}$ \quad Universit\'{e} Paris-Saclay, CNRS/IN2P3, IJCLab,
 91405 Orsay, France; christophe.beigbeder@ijclab.in2p3.fr~(C.B.);
 ion.cojocari@ijclab.in2p3.fr~(I.C.); nicolas.de-sereville@ijclab.in2p3.fr~(N.d.S.); nicolas.dosme@ijclab.in2p3.fr~(N.D.); mariya.georgieva@ijclab.in2p3.fr~(M.G.); clarisse.hamadache@ijclab.in2p3.fr~(C.H.); sally.hankache@ijclab.in2p3.fr~(S.H.); jimmy.jeglot@ijclab.in2p3.fr~(J.J.);\linebreak   beng-yun.ky@ijclab.in2p3.fr~(B.-Y.K.); vincent.lafage@ijclab.in2p3.fr~(V.L.); christine.le-galliard@ijclab.in2p3.fr~(C.L.G.); joseph.mangan@ijclab.in2p3.fr~(J.M.); jean.peyre@ijclab.in2p3.fr~(J.P.); arnaud.saussac@ijclab.in2p3.fr~(A.S.); matias.vecchio@ijclab.in2p3.fr~(M.V.)\\
$^{2}$ \quad School of Physics \& Centre for Space Research, University College Dublin, Dublin 4,  
 Ireland; david.murphy@ucd.ie (D.M.); alexey.uliyanov@ucd.ie~(A.U.); caimin.mckenna@ucdconnect.ie~(C.M.); lorraine.hanlon@ucd.ie~(L.H.); colin.wade1@ucd.ie~(C.W.)\\
$^{3}$ \quad KTH Royal Institute of Technology, Department of Physics, SE-106 91 Stockholm, Sweden; The Oskar Klein Centre for Cosmoparticle Physics, AlbaNova University Center, SE-106 91 Stockholm, Sweden
baranwal@kth.se~(P.B.); mozsi@kth.se~(M.K.); pearce@kth.se~(M.P.); vvarun@kth.se~(V.V.)\\
$^{4}$ \quad AIM-CEA/DRF/Irfu/Département
 d'Astrophysique, CNRS, Université Paris-Saclay, Université Paris Cite, Orme des Merisiers, Bat. 709, 91191 Gif-sur-Yvette, France; arnaud.claret@cea.fr~(A.C.); eric.doumayrou@cea.fr~(E.D.); philippe.laurent@cea.fr~(P.L.); aline.meuris@cea.fr~(A.M.); arjun.poitaya@cea.fr~(A.P.); diana.renaud@cea.fr~(D.R.)}

\corres{Correspondence: nathan.franel@ijclab.in2p3.fr (N.F.); vincent.tatischeff@ijclab.in2p3.fr (V.T.)}

\abstract{COMCUBE-S (Compton Telescope CubeSat Swarm) is a proposed mission aimed at understanding the radiation mechanisms of ultra-relativistic jets from Gamma-Ray Bursts (GRBs). It consists of a swarm of 16U CubeSats carrying a state-of-the-art Compton polarimeter and a {bismuth germanium oxide (BGO)} spectrometer to perform timing, spectroscopic and polarimetric measurements of the prompt emission from GRBs. The mission is currently in a feasibility study phase (Phase A) with the European Space Agency to prepare an in-orbit demonstration. Here, we present the simulation work used to optimise the design and operational concept of the microsatellite constellation, as well as estimate the mission performance in terms of GRB detection rate and polarimetry. We used the MEGAlib software to simulate the response function of the gamma-ray instruments, together with a detailed model for the background particle and radiation fluxes in low-Earth orbit. We also developed a synthetic GRB population model to best estimate the detection rate. These simulations show that COMCUBE-S will detect about 2 GRBs per day, which is significantly higher than that of all past and current GRB missions. Furthermore, simulated performance for linear polarisation measurements shows that COMCUBE-S will be able to uniquely distinguish between competing models of the GRB prompt emission, thereby shedding new light on some of the most fundamental aspects of GRB physics.\vspace{25pt}} 


\begin{document}
\section{Introduction}
\label{sec:introduction}

GRBs are the most luminous electromagnetic explosions in the universe, radiating most of their energy in the form of a burst of gamma-ray photons that lasts from tens of milliseconds to several hundred seconds \citep{piran2004}. The central engines producing the most relativistic jets known in nature, their physical composition, and the processes of energy dissipation and radiation, are all poorly understood, despite more than 50 years having passed since GRBs were discovered \citep{kumar2015}. 

{Due to their extreme luminosity, GRBs can be observed at much greater distances than well-established standard candles in cosmology like Type Ia supernovae}. Empirical correlations---such as the Amati and Ghirlanda relations---offer a link between observable quantities and intrinsic properties, enabling the use of GRBs to trace the Universe's expansion history \cite{amati2013,amati2018}. However, the physical mechanisms driving these correlations remain poorly understood due to limited knowledge of the physics of ultra-relativistic jets. This uncertainty affects our understanding of the correlations' origin, their intrinsic spread, the presence of outliers, and the possibility of evolution with redshift, all of which pose challenges to the reliability of GRBs as cosmological tools \citep{bargiacchi2025}. 

Various GRB emission models predict different levels of linear polarisation in the soft gamma-ray band \citep{toma2009,gill2020}. In models where the GRB jet energy is dominated by magnetic fields, synchrotron radiation in globally ordered magnetic fields produces a highly polarised photon flux, with a typical polarisation degree (PD) of 20--60\%. In matter-dominated models, PD depends on the viewing angle with respect to the jet axis and the size of the portion of the jet visible to the observer (the jet cannot be observed in its entirety because of the high Lorentz factor of the ejecta and the relativistic focusing effect). The main models of GRB prompt emission can be distinguished by measuring the distribution of PD for a representative sample of GRBs, provided that the mission has sufficient polarisation sensitivity to detect more than 60 GRBs with a Minimum Detectable Polarisation (MDP) of $<$30\% \citep{mcconnell2017,pearce2019}. The MDP
 defines the sensitivity of a polarisation measurement in terms of PD. It depends on the properties of both the source and the polarimeter:
\begin{linenomath}
\begin{equation}
MDP = \frac{4.29}{\mu_{100} F_S A_{\rm eff}}\sqrt{F_S A_{\rm eff} + B t},
\end{equation}
\end{linenomath}
where the factor 4.29 applies to a 99\% confidence level for detection, $F_S$ is the source fluence (ph~cm$^{-2}$), $A_{\rm eff}$ the detection effective area (cm$^2$), $B$ the background count rate (s$^{-1}$), $t$ the observing time (s), and $\mu_{100}$ the modulation factor expected for a 100\% polarised source (see, e.g., {Tatischeff et~al.}~\citep{tatischeff2019}).

COMCUBE-S (Compton Telescope CubeSat Swarm) is a proposed mission based on a swarm of 16U CubeSats in low-Earth orbit (LEO) aiming to gain a breakthrough understanding of the fundamental physical mechanism and extreme physical conditions that give rise to the prompt emission of GRBs. It is designed to perform linear polarimetry with sufficient precision to allow discrimination between GRB prompt emission models, as well as low-latency GRB localisations for counterpart observation in the multi-messenger era of time-domain astronomy. 

Here, we present the simulation work carried out during a preliminary study phase with ESA (Phase 0) in order to investigate the GRB detection rate and polarimetric performance of the mission. A brief overview of the mission concept is given in Section~\ref{sec:mission}. The simulation set-up is presented in Section~\ref{sec:simulations} and the GRB population model in Section~\ref{sec:grb_population}. The simulation results are given in Section~\ref{sec:results}, followed by some conclusions in Section~\ref{sec:conclusions}.

\section{COMCUBE-S Mission Concept}
\label{sec:mission}

COMCUBE-S is a swarm of 16U CubeSats that detect GRBs and measure their spectral, temporal and polarisation properties. They also determine GRB positions on the sky and provide fast notifications for follow-up observation by other facilities. Each satellite carries two complementary gamma-ray instruments each with a large field of view, allowing the swarm to offer full-sky coverage on a large spectral band (30~keV--10~MeV). The satellites operate in zenith-pointing mode to provide maximum effective area for sources not occulted by Earth. By simultaneously observing the same GRB and combining the data, the satellites work together as a single distributed telescope with a large effective area. More details on the mission concept are given in \citet{murphy2026}. 

The primary payload on each CubeSat is a Compton telescope optimised for gamma-ray polarimetry. The first scattering stage of the Compton telescope Unit (CTU) consists of eight double-sided silicon strip detectors (DSSDs) spread over two layers, which enable precise measurements of the 3D position of the interaction and energy deposit of each scattering interaction. The scattered gamma-rays are then {absorbed in position-sensitive scintillation detectors placed around the DSSD layers. The absorber layer beneath the DSSDs consists of pixelated gadolinium aluminium gallium garnet (GAGG:Ce) scintillators coupled with OnSemi silicon photomultiplier (SiPM) arrays. The absorber modules around the DSSDs are made of thin (5~mm) monolithic cerium bromide (CeBr$_3$) scintillators coupled with Hamamatsu SiPM arrays.
 Further details on the CTU design are provided in Section~\ref{sec:massmodel}}. The CTU detects both single-interaction events (gamma rays fully absorbed in one detector) in an energy range of 30~keV--2~MeV, which are used to detect GRBs, measure their spectra and light curves, and double-interaction events (Compton scattered gamma-rays) in an energy range of 100~keV--2~MeV for polarisation measurements and Compton imaging. An independent bismuth germanium oxide (BGO) spectrometer extends spectral measurements up to 10 MeV. The BGO Spectrometer Unit (BSU) comprises a BGO crystal of $50 \times 50 \times 150$~mm$^3$, which is read out at both ends by OnSemi SiPM arrays. 
 
 {SiPMs exposed to high proton fluxes are subject to radiation damage that gradually increases their dark current and noise over the mission lifetime~\cite{Mitchell2021, Ulyanov2020, Zheng2022}. This additional noise raises the low-energy detection limit of the scintillation detectors. In COMCUBE-S, the SiPMs will be shielded by at least 3~mm of aluminium, which stops low-energy protons\linebreak   ($<$24 MeV) abundant in the spectra of trapped and solar protons. This shielding substantially reduces, but cannot fully prevent radiation-induced damage of SiPMs. Because the spectra of trapped and solar protons extend to hundreds of MeV, further increases in shielding thickness become progressively less efficient for providing additional protection. The extent of SiPM radiation damage is strongly dependent on the orbital parameters. High-inclination orbits (commonly used in CubeSat missions) are particularly unfavourable, as they expose the spacecraft to trapped protons in the South Atlantic Anomaly\linebreak   (see Section~\ref{sec:nonoperation}) as well as to solar protons at high latitudes. For COMCUBE-S, an equatorial orbit with an altitude of 500~km is considered as the baseline (see Section~\ref{sec:results}), which largely avoids trapped protons and thereby reduces the expected radiation damage by orders of~magnitude.
}

\begin{figure}[H]   
\includegraphics[width=0.8\textwidth]{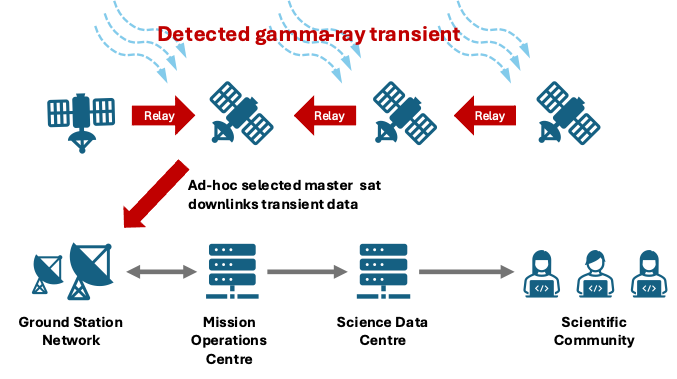}
\caption{Representation of the concept of operations of the COMCUBE-S CubeSat swarm.\label{fig:conops}}
\end{figure}

The concept of operation of the CubeSat swarm is illustrated in Figure~\ref{fig:conops}. The satellites continuously share and combine their observed gamma-ray count rates to detect a simultaneous increase in count rates on multiple satellites, which allows detection of fainter GRBs than is possible by current or proposed missions (see Section~\ref{sec:results}). Once a potential GRB detection is established, all satellites are notified, and a trigger alert period is autonomously established. The satellites form a data link chain and rapidly relay their selected and highly compressed observation data to the satellite in the most favourable position above a ground station (Figure~\ref{fig:conops}). The downlinked data {are} then processed immediately at the Science Data Centre to determine the sky coordinates of the source and other relevant parameters for immediate observation follow-up by ground telescopes and other facilities.

The localisation is first obtained by comparing the count rates in different detectors and on different satellites. The localisation can also be achieved using Compton imaging and by the pseudo-range multilateration method (also called timing triangulation) based on relative time delay measurements, which will increase the reliability and accuracy of the results. {A preliminary analysis based on work carried out for the HERMES constellation concept (see, e.g., \cite{fuschino2019}) shows that using timing triangulation with the COMCUBE-S swarm, a localisation accuracy of better than 1$^\circ$ could be achieved for most long GRBs and around 3$^\circ$ for a significant proportion of short GRBs. More details are given in \citet{murphy2026}.}

{GRB source localisation via triangulation requires timing precision on the order of several tens of microseconds. Distribution of such precise timing information to the individual satellites is in principle easily achievable via a global navigation satellite system (GNSS) receiver which can typically provide a time reference precise to within tens of nanoseconds. Each COMCUBE-S satellite will carry a GNSS receiver for the purposes of timing distribution as well as precise orbit determination which will be used by flight dynamics to plan and verify orbital maneuvers. However, it has been widely demonstrated that in-orbit GNSS receivers can be prone to loss of lock due to both natural \cite{pezzopane2021} as well as synthetic or intentional \cite{wu2024} interference. When such loss of lock occurs, the timing must be sustained onboard by a free-running clock. To maintain high precision, a chip-scale atomic clock may be used. This atomic clock would be disciplined by and aligned to the GNSS timing when available, and continue to provide a precise time standard when GNSS timing is lost.}

While the data {are} processed on ground for localisation, the satellites continue observation of the transient event and send additional data to the ground to refine the localisation. With every localisation, the Science Data Centre {disseminates an alert via the General Coordinates Network (GCN, 
  {\url{https://gcn.nasa.gov/}}) or a similar network, so that it can be quickly acted upon by partnering ground-based facilities for follow-up~observations}.

In addition, each time a GRB is detected, the time-tagged event data recorded by all satellites of the swarm during the burst alert period is transmitted to the ground to enable spectral, temporal and polarisation analyses. Using the whole CubeSat swarm as a large distributed instrument is a significant advantage for the control of systematic errors compared to conventional missions based on a single spacecraft, as each COMCUBE-S satellite observes a given source in the sky in a different geometric configuration.  

\section{COMCUBE-S Performance Simulations}
\label{sec:simulations}

A detailed simulation setup was developed to help define the mission and predict its scientific performance. This enabled us to study, in particular, the required number of satellites in the swarm and the optimal orbit for the targeted scientific objectives.

\subsection{COMCUBE-S Satellite Mass Model}
\label{sec:massmodel}

We used the MEGAlib (Medium-Energy Gamma-ray Astronomy library) software package \cite{zoglauer2006} to simulate the response function of the CubeSats to GRB emission and background radiation. MEGAlib 
 ({\url{https://megalibtoolkit.com/}}) was specifically developed for analysis of simulation and calibration data from hard X-ray/gamma-ray instruments. It includes software tools for geometry and detector description, allowing detailed modelling of different detector types and characteristics. The simulation of the interaction of gamma rays and particles with the satellite (both detector and passive materials) is carried out with the Monte-Carlo simulation package Cosima \cite{zoglauer2009}, which is based on Geant4
\linebreak   ({\url{https://geant4.web.cern.ch/}}). {The detector response is modeled by applying user-defined detector effects to the position and energy deposits simulated with Geant4: spectral and position resolutions, thresholds and efficiencies. Detailed electronics and sensor effects, like the photon-detection efficiency of SiPMs, are included indirectly as a smearing step rather than a detailed photon-by-photon simulation.}
MEGAlib also includes specialised Compton event reconstruction algorithms, which allowed us to simulate the polarisation response of the CTU in detail.

\begin{figure}[H]    
\includegraphics[width=0.85\textwidth]{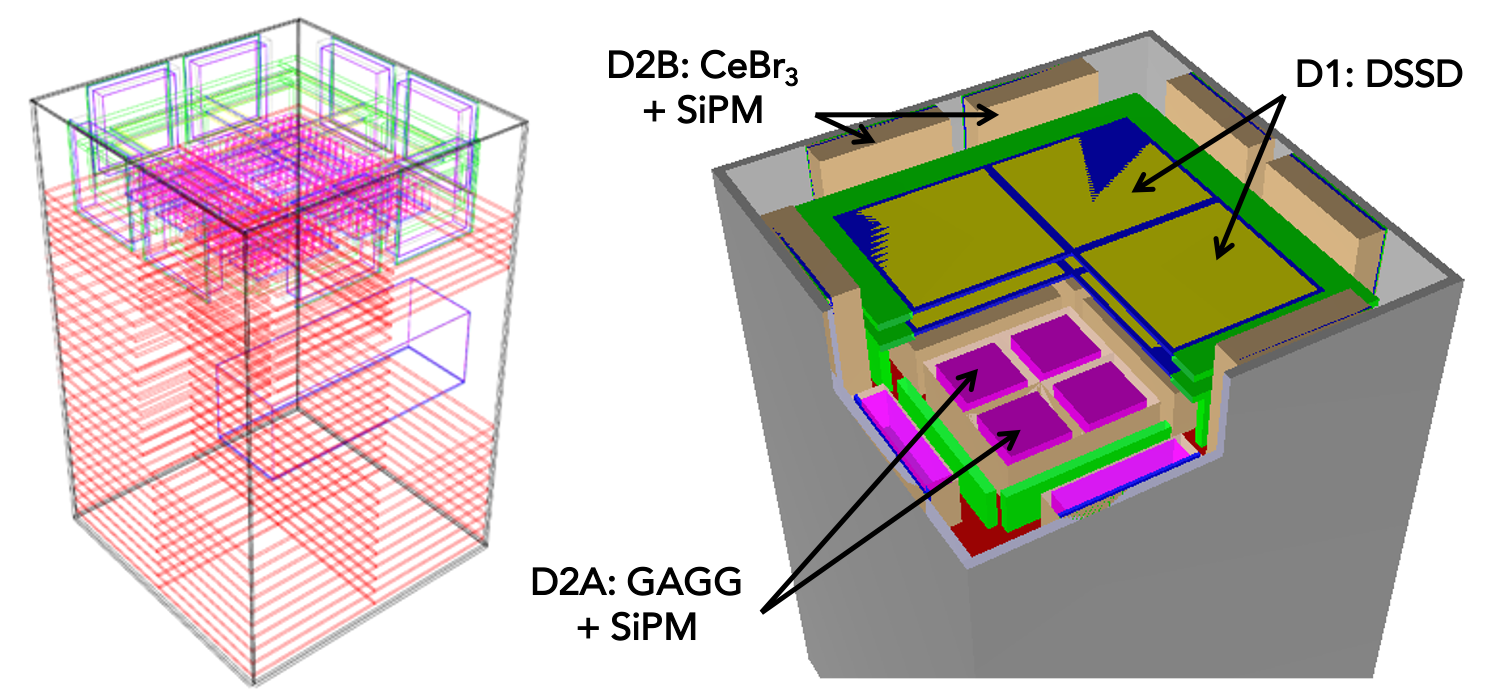}
\caption{COMCUBE-S payload and CubeSat model developed for numerical simulations of the instrument response function with the MEGAlib software.\label{fig:megalib}}
\end{figure}

The Cosima simulations performed for the mission feasibility study used a simplified mass model for the spacecraft and BSU, but a comparatively detailed model for the CTU (Figure~\ref{fig:megalib}). In particular, the detectors of the CTU were modelled based on their precise geometric properties and response---detection threshold, spectral resolution, and position resolution---measured with an instrument prototype in the laboratory. The CTU includes (see Figure~\ref{fig:megalib}): eight DSSDs arranged in two layers (D1), 16 segmented gadolinium aluminium gallium garnet (GAGG:Ce) scintillator detectors (D2A) underneath the DSSDs and eight monolithic cerium bromide (CeBr$_3$) scintillator detectors (D2B) on the sides of the instrument. Simulations showed that the ring of D2B detectors around the perimeter of the D1 scattering layer significantly improves the polarimetric response of the instrument. This is because the Compton scattering cross section for polarimetry is maximised when the scattering angle is close to 90$^\circ$.

\subsection{Background Model}
\label{sec:background}

The background model used in this study is based on the work of \citet{cumani2019} for low-inclination LEOs. It considers: (i) primary cosmic-ray protons, alpha-particles, electrons and positrons, whose intensity is modulated by the Earth's magnetic field, (ii)~secondary protons, neutrons, electrons and positrons produced by cosmic-ray interaction with the Earth's atmosphere, (iii) the Earth's hard X-ray/gamma-ray albedo, and (iv) the cosmic diffuse X- and gamma-ray background. The model of \citet{cumani2019} has been extended to all geomagnetic latitudes, using, in particular, the ISO-15390 cosmic-ray model for the primary proton and alpha-particle fluxes as functions of the geomagnetic position. We used the ApexPy Python class (version 2.1.0) for the geomagnetic-to-geographic coordinate transformation. Further details will be given in \citet{franel2026}. 

\begin{figure}[H]    
\includegraphics[width=0.99\textwidth]{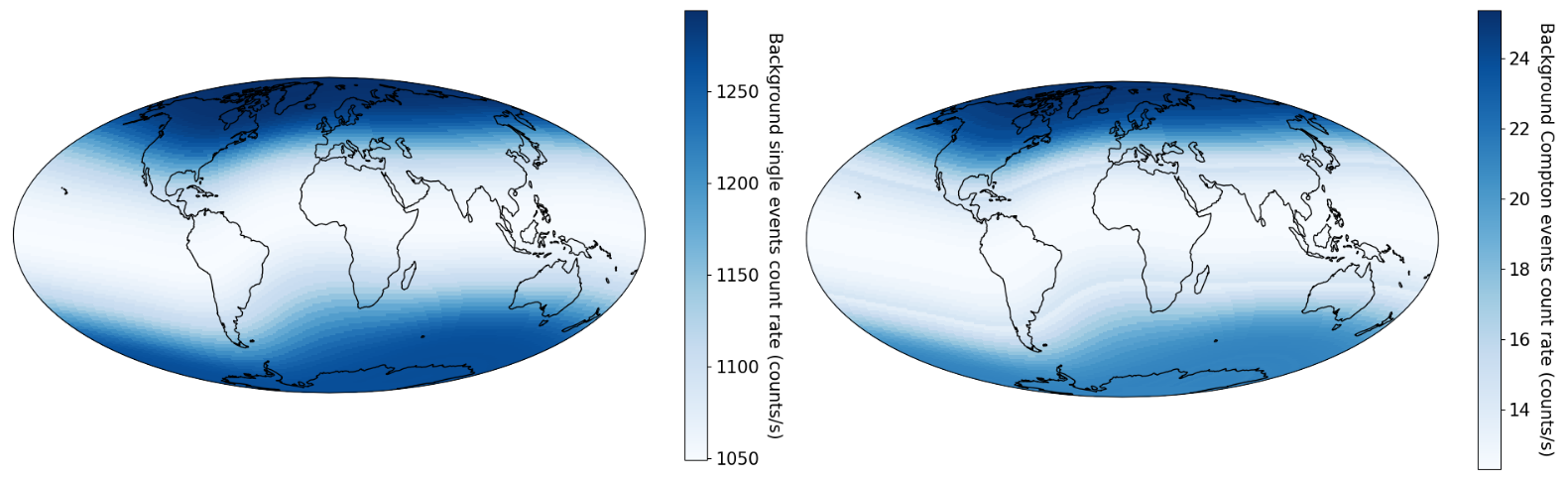}
\caption{Simulated background count rates of the Compton Telescope Unit of a COMCUBE-S satellite for (left) single-interaction events and (right) double-interaction events against geographic position. The satellite is at an altitude of 500~km. The high fluxes of particles trapped inside the inner Van Allen radiation belt in the Earth's polar regions and the South Atlantic Anomaly are not represented in these maps (see Section~\ref{sec:nonoperation}).\label{fig:background}}
\end{figure}

Interaction of background radiation and particles with the COMCUBE-S spacecraft was simulated with MEGAlib. We performed computationally intensive simulations (1.5~days on a 64-core server) to obtain maps of count rates for single- and double-interaction background events as a function of altitude and geographic position (see Figure~\ref{fig:background}). These results were then used to study the scientific performance of the mission for various configurations of the CubeSat swarm. 

{The main background of COMCUBE-S is from cosmic gamma-rays, which contributes 70\% to 80\% depending on the orbit. The second most significant source of background is generated by the Earth's albedo gamma-rays, with a contribution ranging from 10\% to 20\%. The background induced by charged particles and neutrons is less important.
}

\subsection{Non-Operation Area}
\label{sec:nonoperation}

We took into account that the COMCUBE-S satellites will be exposed to strong fluxes of electrons and protons trapped inside the inner Van Allen radiation belt when they will pass through the Earth's polar regions and the South Atlantic Anomaly (SAA). The particle flux can be so intense in these regions that the scientific data collection is generally turned off when a high-energy astronomy satellite passes through them, although some diagnostic data can still be recorded. 

\begin{figure}[H]
\includegraphics[width=0.495\textwidth]{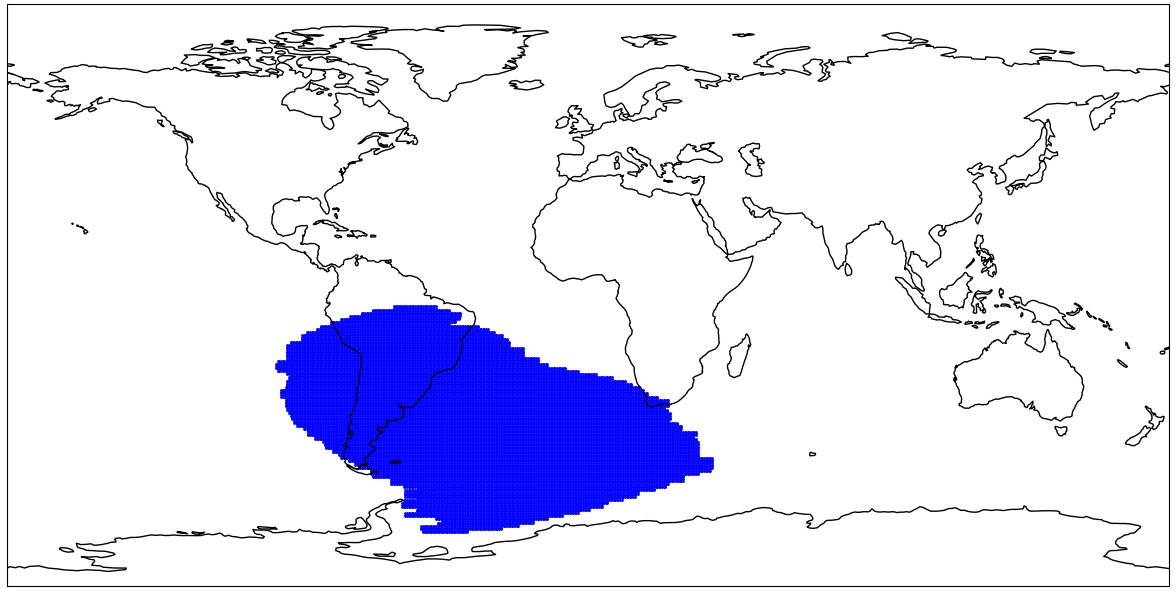}
\includegraphics[width=0.495\textwidth]{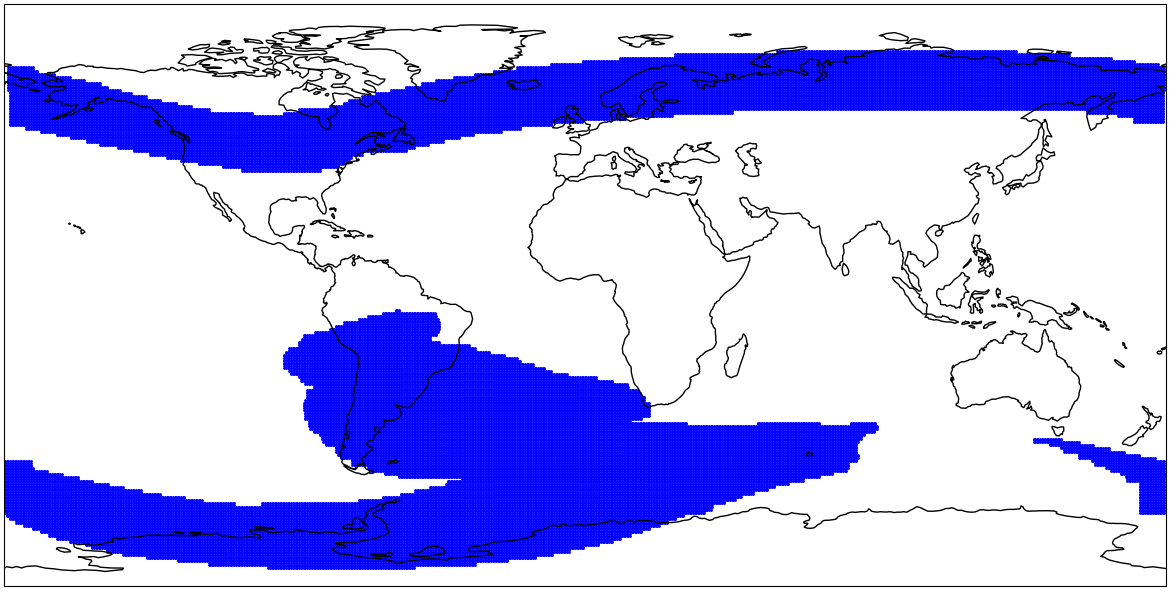}
\caption{Simulated regions of non-observation for COMCUBE-S satellites at 500~km altitude (blue highlighted areas) due to excessive fluxes of trapped protons (left) and electrons (right). Based on NASA's AP8min and AE8max trapped particle models (see text).\label{fig:non-operation}}
\end{figure}

To estimate in which regions the COMCUBE-S satellites will not be operational for gamma-ray sky monitoring, we used NASA's AP8min and AE8max models for trapped protons and electrons, respectively (see
 \url{https://ccmc.gsfc.nasa.gov/models/AE-8_AP-8_RADBELT~1.0}), which have been shown to provide a fair estimate of fluxes measured with the Particle Monitor instrument on-board the BeppoSAX mission \citep{ripa2021}. In the work of \citet{ripa2021}, the more recent IRENE (International Radiation Environment Near Earth) AP9 models were found to severely overestimate the measured fluxes for low inclination orbits. To determine the position of the non-observation zones, which changes over time, we simulated 90 days of polar orbit ($i = 90^\circ$) from 1 January 2030 with a sampling rate of 10 seconds. To find the zone boundaries, we adopted the integral flux limits $F_p~(E_p > 1~{\rm MeV}) = 3$~cm$^{-2}$~s$^{-1}$ and $F_e~(E_e > 1~{\rm MeV}) = 10$~cm$^{-2}$~s$^{-1}$, for protons and electrons respectively, which are about an order of magnitude above the mean particle fluxes outside the radiation belts. Simulated maps are shown in Figure~\ref{fig:non-operation} for satellites orbiting at 500~km altitude. The region of non-observation due to trapped protons (left~panel) corresponds to the {SAA}, while that which is due to trapped electrons (right panel) also includes two wide bands near the poles. 

{The AP8min model predicts that a satellite in an equatorial orbit at an altitude of 500~km will completely avoid the SAA and can therefore achieve a duty cycle of 100\%. But this is not the case according to the AP9 model, which predicts a duty cycle of only about 85\% for an integral flux limit of energetic protons $F_p~(E_p > 1~{\rm MeV}) = 1$~cm$^{-2}$~s$^{-1}$ (see \cite{ripa2021}). However, this should be considered as a worst-case scenario, because the AP9 model significantly overestimates the proton flux for low-inclination orbits. Recently, using SAA measurements from the High-Energy Particle Detector instrument (HEPD-01) aboard the China Seismo-Electromagnetic Satellite (CSES-01), Gallego et~al. \cite{gallego2025} estimated that the AP9 model overestimates the proton flux in the SAA by nearly an order of magnitude for an altitude of 500 km and a latitude in the range [$-1^\circ$; $+1^\circ$]. Taking into account a duty cycle $>85$\% for an equatorial orbit would have a limited impact on the estimated performance of the COMCUBE-S CubeSat swarm. For orbits with inclination $i = 45^\circ$ or $97.5^\circ$ (see Table~\ref{tab:setup} below), the AP8min and AP9 models are in relatively good agreement. 
  }

\begin{table}[H] 
\caption{COMCUBE-S swarm configurations studied.\label{tab:setup}}
\centering
\begin{tabularx}{\textwidth}{CCC}
\toprule
\textbf{Number of Spacecraft} & \textbf{Orbit Inclination} & \textbf{Orbit Altitude} \\
\midrule
27 & $0^\circ$, $45^\circ$ and $97.5^\circ$ & 400~km and 500~km \\
36 & $0^\circ$, $45^\circ$ and $97.5^\circ$ & 500~km \\
\bottomrule
\end{tabularx}
\end{table}

{Trapped protons interacting with a spacecraft can produce various radioisotopes, whose decay can contribute to the background once the satellite has left the SAA and resumed observations (see \cite{weidenspointner2001} for the modeled background of the CGRO/COMPTEL instrument). \citet{cumani2019} studied this background component using MEGAlib simulations and showed that it should generally be negligible for a gamma-ray satellite in a LEO with inclination $i < 10^\circ$ (see also Ref.~\cite{tatischeff2022}). However, for higher-inclination orbits, the background caused by the decay of relatively short-lived radioisotopes can limit observations for a few minutes after leaving the SAA. Taking into account such a cool-down period after each SAA passage would have no significant impact on the performance of the COMCUBE-S swarm.
}

{Intense particle fluxes in the inner Van Allen belt can also cause afterglow emission in scintillation crystals, i.e., residual, long-lasting phosphorescence emission. This effect could also limit observations just after leaving the trapped particle regions. BGO and CeBr$_3$ scintillation crystals are known to exhibit very low and brief afterglow, decaying in a few ms~\cite{Yanagida2014, Koppert2019}. However, GAGG crystals exhibit relatively intense phosphorescence for several minutes after they have been exposed to ionizing radiation~\cite{Lucchini2016, Dilillo2022}. This slow emission leads to an increase in the detector current and noise. In a study for the Fast Gamma-ray Spectrometer~\cite{Pallu2024}, a $2\times 2\times 1$ cm$^3$ GAGG crystal was irradiated for 15 min by 30--70 MeV protons with a flux of 10,000 particles~cm$^{-2}$~s$^{-1}$ to simulate a radiation exposure during one SAA passage. The background spectra acquired immediately after the irradiation showed an increased rate of noise counts below 25 keV which were attributed to GAGG phosphorescence. COMCUBE-S uses GAGG crystals that are approximately 7~times smaller in volume, which proportionally reduces the intensity of phosphorescence per crystal. Assuming the magnitude of the detector noise is proportional to the square root of the average emission intensity, the noise of COMCUBE-S detectors associated with GAGG phosphorescence is not expected to exceed 10~keV and will be suppressed by the detector threshold. In the worst case scenario, the CubeSat may need to be excluded from GRB observation by the COMCUBE-S swarm for a few additional minutes after it passes through the SAA, which would result in negligible impact on the swarm performance.
}

\subsection{Simulation Setup}
\label{sec:setup}

All 16U CubeSats will be placed on a single orbit to maximize inter-satellite link speed and minimize GRB alert latency for ground-based facility follow-ups. A single orbit also greatly simplifies operation planning, orbit maintenance and automation and maximizes ground station usage efficiency. We considered two orbit altitudes in this study, 400~km and 500~km, and three inclinations, $0^\circ$, $45^\circ$, and $97.5^\circ$, the latter inclination being representative of a Sun-synchronous orbit (SSO) at 500~km altitude. The spacecraft number should be optimised for the required scientific measurements, from a trade-off between scientific performance and cost. We considered 27 and 36 satellites in the study. The various swarm configurations studied are summarised in Table~\ref{tab:setup}.

GRBs are simulated to occur at a random time and random position in the sky. When a GRB occurs, the position of each satellite in the orbit is calculated and the payload response to the incident GRB photon flux is simulated with MEGAlib (see Section~\ref{sec:massmodel}). However, simulations are not carried out for satellites that are in regions of non-observation at the time of the burst (see Section~\ref{sec:nonoperation}) or when the GRB is occulted by the Earth. 

To simulate the GRB fluences, peak fluxes and spectra, we first used the Fourth {\em Fermi}-GBM Gamma-Ray Burst Catalogue, which covers the first 10 years of operation of the {\em Fermi} mission \cite{vonkienlin2020}. We selected the GRBs for which the total emission spectrum has been fitted by a theoretical model (see \cite{Poolakkil2021}), resulting in 1934 long and 367 short GRBs. Four best-fit models of emission spectra are considered in the catalogue: a power law, an exponentially attenuated power-law, a smoothly broken power-law, and a Band function \cite{Band1993}. The GRB light curves are not directly available in the {\em Fermi}-GBM catalogue. They were reconstructed for each GRB from the {\em Fermi}-GBM online database containing the Time-Tagged Events
(see \url{https://fermi.gsfc.nasa.gov/ssc/data/access/gbm/}). Considering actual GRB light curves instead of mean fluxes and $T_{90}$ values---the interval containing 90\% of the burst’s detected photons---is important for time-resolved polarisation analyses and for testing different trigger strategies.


Simulations performed using the {\em Fermi}-GBM catalogue showed that COMCUBE-S would detect virtually all GRBs in the catalogue, which reflects the greater sensitivity of the CubeSat swarm mission. The few GRBs not detected by COMCUBE-S at each random draw are very faint events randomly positioned in low-sensitivity areas of the sky. While the {\em Fermi}-GBM catalogue is ideal for studying the mission performance for GRB polarimetry, the detection performance should be evaluated using a population of synthetic GRBs, resembling those detected by {\em Fermi}-GBM at high fluxes, and extending to fainter fluxes.

\section{Synthetic GRB Population Model}
\label{sec:grb_population}

The synthetic GRB population model is based on empirical distributions of GRB spectral, timing and luminosity parameters. Due to their different nature, short and long bursts were described using different parameters. We used a Band model to simulate the GRB spectra, with spectral indices randomly drawn from distributions extracted from the {\em Fermi}-GBM catalogue \cite{Poolakkil2021}. GRB duration was also obtained from the {\em Fermi}-GBM catalogue, from a random sampling of the measured $T_{90}$ distribution. However, in order to simulate light curves similar to those detected, we assigned to each synthetic GRB a light curve based on the one of the GBM burst with the closest $T_{90}$ value. Each light curve was then slightly stretched or compressed in time to exactly match the randomly drawn $T_{90}$ duration.

The isotropic rest-frame peak luminosity distributions were modeled as broken power-laws, taking the spectral indices and the break luminosity as free parameters. We use the slightly modified versions of the Yonetoku empirical relation~\cite{yonetoku2004} given in \mbox{\citet{Ghirlanda2016}} and~\citet{Lien2014} to derive the rest-frame peak energy as a function of the rest-frame peak luminosity. The GRB rates as a function of redshift were also taken from~\citet{Ghirlanda2016} and \citet{Lien2014} for short and long GRBs, respectively. Each redshift distribution has four free parameters: the current GRB rate per Gpc$^3$, the redshift at the peak of the distribution and two parameters describing the shape of the distribution around the peak. The adopted cosmological model is a flat $\Lambda$ Cold Dark Matter ($\Lambda$CDM) model with a Hubble constant $H_0 = 70$~km~s$^{-1}$~Mpc$^{-1}$ and a matter density at $z = 0$ $\Omega_{m,0} = 0.3$.

Extensive Monte Carlo simulations were used to fit the synthetic GRB populations to the {\em Fermi}-GBM data sets for both short and long GRBs. The fits were performed to the measured peak flux distributions using $\chi^2$ statistic. We assumed that {\em Fermi}-GBM detects all GRBs with peak flux $F_{\rm peak} \ge F_{\rm comp}=10$~ph~cm$^{-2}$~s$^{-1}$ when they are in the field of view of the instrument (fraction of sky coverage $\Omega_{\rm GBM} = 0.69$). We also took into account the duty cycle of GBM:  $f_{\rm GBM} = 0.85$. The adopted completeness limit $F_{\rm comp}$ is more than an order of magnitude higher than the trigger threshold, $F_{\rm trig}=0.74$~ph~cm$^{-2}$~s$^{-1}$ \citep{meegan2009}. The model has seven free parameters for each GRB population: three parameters describing the peak luminosity of the GRBs and four describing their redshift distribution. These free parameters were randomly drawn from uniform distributions within reasonable ranges taken from the literature \cite{Ghirlanda2016}. The best-fit parameters were found to be in good agreement with previous works. Further details will be given in \citet{franel2026}. The synthetic GRB catalogue predicts that an ideal GRB mission covering the entire sky with perfect detection sensitivity would detect up to 1570 short GRBs and 5503 long GRBs each year.

\begin{figure}[H]    
\includegraphics[width=0.99\textwidth]{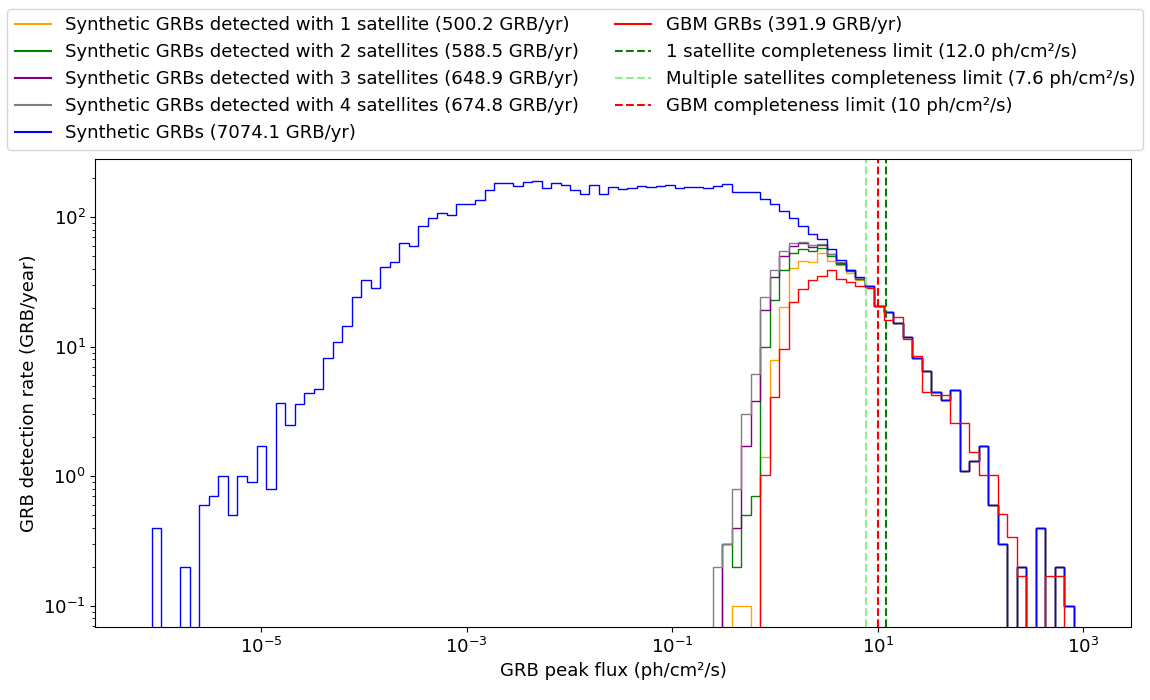}
\caption{Number of GRBs detected per year as a function of peak flux for different GRB populations. The blue histogram shows the total population of GRBs in the Universe that would be detected by an ideal instrument. The red histogram shows the distribution of GRBs in the {\em Fermi}-GBM catalogue corrected for the field of view and duty cycle of the GBM instrument ($\Omega_{\rm GBM}f_{\rm GBM}=0.59$; see text). The other four histograms show the synthetic GRBs detected by COMCUBE-S, assuming between one and four satellites involved in the trigger scheme (see Section~\ref{sec:results}). Results for COMCUBE-S were obtained assuming a constellation of 27 satellites in an equatorial orbit at an altitude of 500~km. The vertical hatched lines show the completeness limits of the {\em Fermi}-GBM and synthetic catalogues with respect to the simulated GRB total population. \label{fig:grb_population}}
\end{figure}

Figure \ref{fig:grb_population} shows the peak flux distributions for the total population of GRBs in the {\em Fermi}-GBM catalogue and for the synthetic GRB population, in red and blue, respectively. The other histograms show the synthetic GRBs detected by COMCUBE-S for various trigger conditions, as simulated (see Section~\ref{sec:results}). We see that the synthetic GRB population includes GRBs with peak fluxes extending below the minimum flux given in the {\em Fermi}-GBM catalogue, while being in excellent agreement with the detected distribution above $10$~ph~cm$^{-2}$~s$^{-1}$. This makes it a realistic GRB population that can be used to accurately estimate the detection performance of COMCUBE-S.

\section{COMCUBE-S Simulation Results}
\label{sec:results}

Extensive simulations of the COMCUBE-S mission performance were performed for the various swarm configurations shown in Table~\ref{tab:setup}, using the synthetic GRB population to predict the GRB detection rate (Section~\ref{sec:grbrate}) and the {\em Fermi}-GBM catalogue for the polarimetric performance (Section~\ref{sec:polarperf}).

\subsection{GRB Detection Rate}
\label{sec:grbrate}

On-board software on each COMCUBE-S satellite will constantly monitor the detector count rates in a time window of a certain size (trigger timescale) while comparing it to the average background rate calculated from a comparatively long preceding time interval. To efficiently detect GRBs with different timing properties, the burst detection algorithm will typically implement nine trigger timescales ranging from 16~ms to 4.096~s. As any GRB should be observed by multiple satellites of the swarm, requiring a simultaneous count rate increase on several satellites can be used to reduce the trigger threshold and detect fainter GRBs. Table~\ref{tab:triggerthreshold} shows the trigger thresholds required to suppress the rate of false triggers from background fluctuations to a level of less than 0.03 global triggers per day (i.e., simultaneous detections by multiple satellites) for each trigger timescale. These thresholds ensure that the total rate of false triggers across all timescales remains below 0.27 global triggers per day, which is negligible compared to the expected rate of genuine GRB triggers of about two per day (see below).


\begin{table}[H]
\small 
\caption{Trigger thresholds in units of the standard deviation of the background count rate as functions of the trigger timescales and the number of satellites triggering simultaneously. \label{tab:triggerthreshold}}
\centering
\begin{tabularx}{\textwidth}{CC|CCCCCCCCC}
\toprule
\multicolumn{2}{c|}{\textbf{Number of}} & \multicolumn{9}{c}{\textbf{Trigger Timescale in s}}  \\
\multicolumn{2}{c|}{\textbf{Coincident Triggers}}  & \textbf{0.016} & \textbf{0.032} & \textbf{0.064} & \textbf{0.128} & \textbf{0.256} & \textbf{0.512} & \textbf{1.024} & \textbf{2.048} & \textbf{4.096} \\
\midrule
    \multicolumn{2}{c|}{\textbf{1}}   & 7.8   & 7.1   & 6.8   & 6.5   & 6.2   & 5.9   & 5.8   & 5.6   & 5.4   \\
    \multicolumn{2}{c|}{\textbf{2}}   & 5.4   & 5.1   & 4.9   & 4.7   & 4.5   & 4.4   & 4.3   & 4.2   & 4.1   \\
    \multicolumn{2}{c|}{\textbf{3}}   & 4.4   & 4.2   & 4.1   & 3.9   & 3.8   & 3.7   & 3.6   & 3.5   & 3.4   \\
    \multicolumn{2}{c|}{\textbf{4}}   & 4.0   & 3.7   & 3.6   & 3.5   & 3.4   & 3.3   & 3.3   & 3.2   & 3.1   \\
\bottomrule
\end{tabularx}
\end{table}

Table~\ref{tab:grbrate} shows the predicted GRB detection rates of COMCUBE-S for one to four satellites triggering a detection in coincidence. The GRB detection rates were simulated from the synthetic populations of short and long GRBs (Section~\ref{sec:grb_population}). The ranges of detection rates given in Table~\ref{tab:grbrate} correspond to different populations of synthetic GRBs that fit the {\em Fermi}-GBM catalogue for the brightest bursts equally well ($\chi^2 \le \chi^2_{\rm min}+1$). We see that with four triggering satellites the predicted detection rate amounts to more  than 595~GRB~yr$^{-1}$. Even using just one satellite, the minimum rate is predicted to be 444~GRB~yr$^{-1}$, which significantly exceeds the {\em Fermi}-GBM detection rate of 230~GRB~yr$^{-1}$ \cite{vonkienlin2020}. In fact, COMCUBE-S is predicted to detect GRBs at a rate significantly higher than that of all past and current GRB missions. These detection performances are enabled not only by the capabilities of each instrument, but also by the increased likelihood of favourable observation setups when several satellites are distributed around the Earth.

\begin{table}[H] 
\small 
\caption{Annual detection rates of GRBs by COMCUBE-S for a swarm of 27 satellites in an equatorial orbit at an altitude of 500~km. Results are given for different number of satellites triggering simultaneously. Detection-rate ranges reflect different synthetic GRB populations that reproduce the {\em Fermi}-GBM bright-burst sample ($F_{\rm peak} \ge F_{\rm comp}$; see Section~\ref{sec:grb_population}) equally well. \label{tab:grbrate}}
\centering
\begin{tabularx}{\textwidth}{C>{\centering\arraybackslash}m{2cm}CCCCC}
\toprule
\multicolumn{2}{c}{\textbf{Number of Satellites Triggering in Coincidence}} & \textbf{1} & \textbf{2} & \textbf{3} & \textbf{4} \\
\midrule
    \multirow{3}{*}{\textbf{GRB detection rate (GRB/yr)}}                      & \textbf{All GRBs}   & $444$--$622$ & $520$--$733$ & $573$--$806$ & $595$--$838$ \\
                                                         & \textbf{Short GRBs} & $96$--$136$   & $120$--$171$   & $137$--$196$   & $144$--$207$   \\
                                                         & \textbf{Long GRBs}  & $334$--$506$ & $382$--$589$ & $414$--$643$ & $429$--$667$ \\
\bottomrule
\end{tabularx}
\end{table}

\subsection{Polarimetric Performance}
\label{sec:polarperf}

COMCUBE-S provides a new means of measuring gamma-ray polarisation by using the whole CubeSat swarm as a large distributed polarimeter. As each satellite of the constellation views a given source in the sky in a different geometric configuration, it measures a polarigram (distribution of azimuthal Compton scattering angles) with a different modulation factor. The results must then be combined to obtain the polarimetric properties of the source in a common reference frame (in practice, we use the geocentric equatorial coordinate system in our calculations). The combination of geometry-corrected polarigrams provides a means of regularly checking and updating the response functions of the instruments, based on a statistical comparison of the individual results obtained with each CubeSat. This represents a significant advantage for COMCUBE-S over a conventional mission based on a single polarimeter, for which such in-flight updating of the gamma-ray polarisation response is not possible. 

{A Stokes-based analysis~\cite{Kislat2015} can be used for combining data from different CubeSats, whereby a weight is applied to each event to account for the modulation factor, which depends on the orientation of the satellite relative to the burst, the photon energy, etc. The use of Stokes parameters eliminates the need for angular binning (as required for polarigrams) and their additive nature also simplifies the treatment of measurement backgrounds. Once Stokes parameters are determined, a Bayesian analysis (see examples in Refs.~\cite{Chauvin2017, Abarr2020, Kiss&Pearce2024}) can be used to take into account the measurement bias resulting from PD being a positive-definite quantity~\cite{Quinn2012, Maier2014}. In the event of limited statistics, especially on a satellite-by-satellite level, the relative measurement bias due to uncertainties, e.g., in the modulation factor, can become more severe~\cite{Mikhalev2018}. Such cases can therefore benefit from a maximum-likelihood-based analysis as described in Tomsick et~al.~\cite{Tomsick2022}.}

\begin{figure}[H]    
\includegraphics[width=0.8\textwidth]{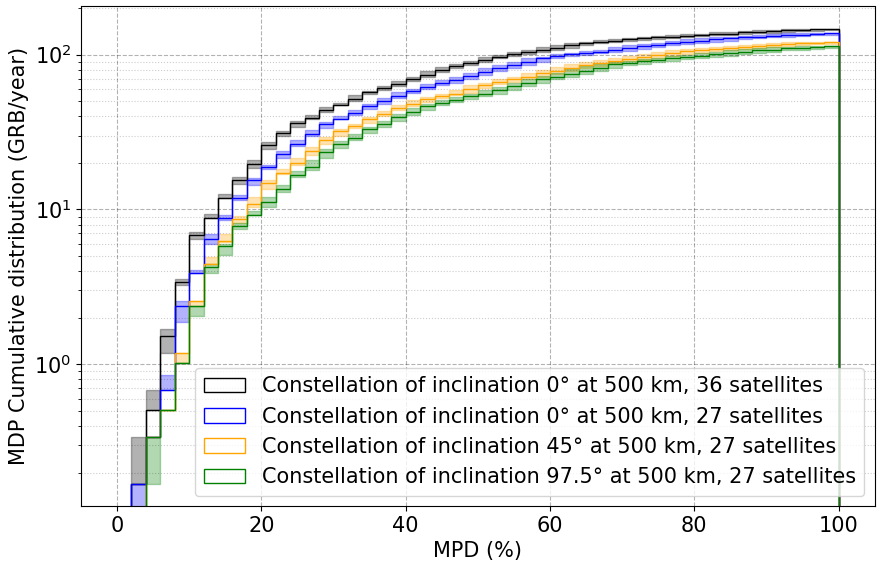}
\caption{Number of GRBs detected per year as a function of their Minimum Detectable Polarisation in the energy range 100--460~keV for various swarm configurations.\label{fig:mdp}}
\end{figure}

{Polarisation measurements will only be possible for the brightest GRBs. Simulations have shown that a minimum GRB fluence of about 20 photons~cm$^{-2}$ between 100~keV and 460~keV will be necessary. We used the {\em Fermi}-GBM catalogue for these simulations (see Section~\ref{sec:setup}), as it is expected to be complete for such bright GRBs}. Figure~\ref{fig:mdp} shows the cumulative distributions of MDP obtained from simulations of various COMCUBE-S swarms in orbits at 500~km altitude. Polarimetric performance is reported for the\linebreak   100--460~keV energy range, which excludes the strong background line at 511~keV. Analyses extending the range up to 1~MeV yield slightly weaker performance (higher MDP) than those limited to 460~keV. The Figure shows that swarms in equatorial orbits (or with more satellites) result in more GRBs being detected with a low MDP, illustrating the fact that polarimetric measurements of weakly polarised GRBs are more likely to be feasible.

Table \ref{tab:mdp} presents the expected detection rate of GRBs with an MDP of at most 10$\%$,\linebreak   30$\%$ and 50$\%$ for various swarm configurations. {Systematic errors on the polarisation include uncertainties in modulation, GRB localisation and satellite pointing precision. However, given the mission requirement that the satellite pointing knowledge shall be known with an accuracy of less than 0.5$^\circ$, the latter error can be safely neglected. Simulations showed that the error in GRB localisation will not significantly affect the polarisation measurements if the GRB position is determined with an accuracy better than 5$^\circ$ (at the 95\% confidence level) in the equivalent radius of the localisation region. The source localisation for polarimetry will be obtained by ground processing, combining the three methods available for the CubeSat swarm: Compton imaging, time triangulation and the relative detector count rate method (Section~\ref{sec:mission}). Furthermore, it is likely that bright GRBs suitable for polarisation measurements will also be identified and localised by other astronomical~observatories.}

The main scientific requirement of the mission is to detect more than 60~GRBs with an MDP of at most 30$\%$, to be able to distinguish between the various models of GRB prompt emission (see Section~\ref{sec:introduction}). Equatorial constellations demonstrate the best capabilities. We see that a configuration of 27 satellites at an altitude of 500 km would meet the main scientific requirement for the polarisation measurements after two years of nominal mission operation in orbit. The other studied configurations for higher-inclination orbits are less favorable, because of a higher background rate (see Figure~\ref{fig:background}) and a lower duty cycle due to the satellite passing through the non-observation regions associated with the inner radiation belt (see Figure~\ref{fig:non-operation}). The main objectives of COMCUBE-S could still be achieved with non-equatorial orbits, but at the cost of a larger number of CubeSats or a longer mission duration in orbit.

\begin{table}[H] 
\small 
\caption{Expected detection rate of GRBs with an MDP of at most 10$\%$, 30$\%$ and 50$\%$. The analysis considered different constellations of 27 or 36 satellites with orbital inclinations of 0°, 45° or 97.5° at altitudes of 400~km or 500~km. The values in blue are the ones compatible with the scientific requirement of the mission to detect more than 60 GRBs with an MDP of at most 30$\%$, assuming a minimum nominal in-orbit lifetime of two years.\label{tab:mdp}}
\centering
\begin{tabularx}{\textwidth}{CCCCCCC}
\toprule
\multicolumn{3}{c}{\textbf{Orbital Configuration}} & \multicolumn{3}{c}{\textbf{GRB Detection Rate (GRB/yr)}} \\
\midrule
\textbf{Altitude}	& \textbf{Orbit Inclination}	& \textbf{Number of Satellites} & \textbf{MDP \boldmath{$\leq$} 10$\%$}	& \textbf{MDP \boldmath{$\leq$} 30$\%$}	& \textbf{MDP \boldmath{$\leq$} 50$\%$}\\
\midrule
   500~km & 0°             & 36 & $3.4^{+0.2}_{-0.2}$  & \textcolor{blue}{$43.8^{+1.9}_{-1.4}$}  & $88.7^{+2.0}_{-2.9}$  \\
    500~km & 0°             & 27 & $2.4^{+0.2}_{-0.5}$  & \textcolor{blue}{$35.6^{+1.2}_{-1.8}$}  & $72.7^{+3.4}_{-3.1}$  \\
    500~km & 45°            & 27 & $1.2^{+0}_{-0.2}$    & $27.9^{+1.5}_{-1.4}$  & $59.9^{+3.6}_{-3.2}$  \\
    500~km & 97.5°          & 27 & $1.0^{+0}_{-0}$      & $23.3^{+1.4}_{-1.7}$  & $53.6^{+2.0}_{-2.0}$  \\
    400~km & 0°             & 27 & $2.0^{+0.7}_{-0}$    & \textcolor{blue}{$34.1^{+1.0}_{-0.7}$}  & $70.5^{+3.9}_{-2.4}$  \\
    400~km & 45°            & 27 & $1.7^{+0}_{-0}$      & $29.3^{+1.2}_{-1.5}$  & $62.5^{+1.9}_{-3.2}$  \\
    400~km & 97.5°          & 27 & $1.2^{+0}_{-0}$      & $24.2^{+1.5}_{-0.7}$  & $55.3^{+3.6}_{-1.7}$  \\
\bottomrule
\end{tabularx}
\end{table}

\section{Conclusions}
\label{sec:conclusions}

The COMCUBE-S CubeSat swarm will surpass conventional GRB missions in both gamma-ray polarimetry and the rapid dissemination of alerts for multi-wavelength follow-up, typically within one to a few minutes. Unlike a single space telescope, the swarm will provide continuous, all-sky coverage, offering a decisive advantage for real-time GRB detection and monitoring. By functioning as a large, distributed aperture, the swarm will enable high precision polarimetric, spectroscopic, and timing observations, allowing for the most comprehensive studies to date on a large sample of GRBs. 

Our detailed design simulations have shown that the scientific requirements of the mission can be met after two years of nominal operation of a swarm of 27 16U CubeSats in equatorial LEO at an altitude of 500 km. {While the science requirements can be satisfied in two years, the mission architecture would support operations beyond this nominal duration}.~The mission will enable competing models of GRB prompt emission to be unambiguously discriminated by uniquely integrating fine polarisation measurements with high-resolution timing and spectroscopy. This comprehensive approach will provide new insights into the strength and structure of magnetic fields in GRB jets, the underlying radiation mechanisms, as well as the effect of the viewing angle – shedding new light on some of the most fundamental aspects of GRB physics. 

In addition to its breakthrough measurements of GRB polarisation across a large sample, COMCUBE-S will play a key role in the emerging field of multi-messenger astrophysics, operating in concert with gravitational-wave and neutrino observatories. Furthermore, COMCUBE-S will be capable of detecting other types of high-energy transient sources than GRBs, both galactic and extra-galactic, and will therefore have a significant impact more broadly in the field of time-domain astronomy. {COMCUBE-S will record data for observatory science of a variety of sources emitting soft gamma-rays, including microquasars, pulsars, blazars, solar flares and terrestrial gamma-ray flashes. These data will be made available to the scientific community to enable join analyses of the timing, spectral and polarisation properties of these sources.}



\vspace{6pt} 

\authorcontributions{Conceptualization, N.F., V.T., D.M., A.U., C.M. and L.H.; methodology, N.F., V.T., D.M., A.U., C.M. and L.H.; software, N.F., A.U. and C.M.; validation, N.F., V.T., D.M., A.U., C.M. and L.H.; writing---original draft preparation, V.T., N.F. and D.M.; writing---review and editing, N.F., V.T., D.M., A.U., C.M., L.H., P.B., C.B., A.C., I.C., N.d.S., N.D., E.D., M.G., C.H., S.H., J.J., M.K., B.-Y.K., V.L., P.L., C.L.G., J.M., A.M., M.P., J.P., A.P., D.R., A.S., V.V., M.V. and C.W. All authors have read and agreed to the published version of the manuscript.}

\funding{This research was funded by the European Space Agency contract No. 4000142483/23/NL/GLC/ov for COMCUBE-S pre-Phase A activities related to the OSIP campaign ‘Innovative Mission Concepts Enabled by Swarms of CubeSats’. N.F. acknowledges financial support from CNES (Centre National d’Études Spatiales) and CNRS (Centre National de la Recherche Scientifique) through contract CNES No. 5100020274 / CNRS No. 260616. LH acknowledges Research Ireland grant 19/FFP/6777. CMK acknowledges Research Ireland grant GOIPG/2024/3488}.

\dataavailability{The simulation software used to support the conclusions of this article will be made available by the authors upon request.} 

\acknowledgments{We would like to express our sincere gratitude to Zsolt Varhegyi, Phil Webster, Derek Bennet and all the staff at AAC Clyde Space involved in the development of the COMCUBE-S~mission.}

\conflictsofinterest{The authors declare that the research was conducted in the absence of any commercial or financial relationships that could be construed as a potential conflict of interest.} 

\abbreviations{Abbreviations}{
The following abbreviations are used in this manuscript:
\\

\noindent 
\begin{tabular}{@{}ll}
BSU & BGO Spectrometer Unit\\
CTU & Compton Telescope Unit\\
GBM & Gamma-ray Burst Monitor\\
GCN & General Coordinates Network\\
GRB & Gamma-ray burst\\
LEO & Low-Earth orbit\\
MDP & Minimum detectable polarisation\\
PD & Polarisation degree\\
SAA & South Atlantic Anomaly\\
SiPM & Silicon photo-multiplier\\
SSO & Sun-synchronous orbit
\end{tabular}
}

\begin{adjustwidth}{-\extralength}{0cm}

\reftitle{References}



 
\isAPAandChicago{}{%

}{}
\PublishersNote{}
\end{adjustwidth}
\end{document}